\documentclass[aps,prc,twocolumn,noshowpacs, superscriptaddress,floatfix]{revtex4}

\usepackage{amsmath}
\usepackage{amssymb}
\usepackage{CJK}
\usepackage{graphicx}
\usepackage{dcolumn}
\usepackage{bm}
\usepackage{color}
\usepackage{ulem}

\begin{document}

\begin{CJK*}{GBK}{song}
\title{Mean free path and shear viscosity in central $^{129}$Xe+$^{119}$Sn collisions below 100 MeV/nucleon}

\author{H. L. Liu}
\affiliation{Shanghai Institute of Applied Physics, Chinese Academy of Sciences, Shanghai 201800, China}
\affiliation{University of the Chinese Academy of Sciences, Beijing 100080, China}
\affiliation{School of Physical Science and Technology, ShanghaiTech University, Shanghai 201203, China}
\author{Y. G. Ma}\thanks{Email: Corresponding author. ygma@sinap.ac.cn}
\affiliation{Shanghai Institute of Applied Physics, Chinese Academy of Sciences, Shanghai 201800, China}
\affiliation{School of Physical Science and Technology, ShanghaiTech University, Shanghai 201203, China}
\author{A. Bonasera}
\affiliation{Cyclotron Laboratory, Texas A$\&$M University, College Station, TX 77843, USA}
\affiliation{INFN, Laboratori Nazionali del Sud, Catania, Italy}
\author{X. G. Deng}
\affiliation{Shanghai Institute of Applied Physics, Chinese Academy of Sciences, Shanghai 201800, China}
\affiliation{University of the Chinese Academy of Sciences, Beijing 100080, China}
\author{O. Lopez}
\affiliation{Laboratoire de Physique Corpusculaire, ENSICAEN, Universi\'te de Caen Basse Normandie, CNRS/IN2P3, F-14050 Caen Cedex, France}
\author{M. Veselsk$\acute{y}$}
\affiliation{Institute of Physics, Slovak Academy of Sciences, D$\acute{u}$bravsk$\acute{a}$ cesta 9, 84511 Bratislava, Slovakia}

\date{\today}

\begin{abstract}
Thermal and transport properties of hot nuclear matter formed in central $^{129}$Xe + $^{119}$Sn collisions at the  Fermi energy  are investigated using the isospin-dependent quantum molecular dynamical (IQMD) model. Temperature ($T$), average density ($\rho$), chemical potential ($\mu$), mean momentum ($P$),  shear viscosity ($\eta$) and entropy density ($s$) are obtained from the phase-space information. The mean free path ($\lambda_{nn}$) and the in-medium nucleon-nucleon cross section ($\sigma_{nn}$) in the highest compressible stage at different incident energies are deduced and compared with the experimental results from Phys. Rev. C $\bf{90}$ (2014) 064602. The result shows that $\lambda_{nn}$ and $\sigma_{nn}$ have the same trend and similar values as the experimental results when the beam  energy is greater than 40 MeV/u at maximum compressed state. Furthermore, the derived shear viscosity over entropy density ($\eta/s$) shows a decreasing behaviour to a saturated value around $\frac{3}{4\pi}$ as a function of incident energy.
\end{abstract}


\maketitle

\section{Introduction}
\label{introduction}
In the past decades, extensive experimental and theoretical efforts have been devoted to investigating the properties of nuclear matter with the help of heavy ion collision (HIC)~\cite{samb2001-1,amcp2000-2,ygm2004-3,mlg1994-4,bmf2008-5,jpo1995-6,ygm1997-7,ygm1999-8,ygm2002-9}. Many probes have been suggested to be linked to the properties of nuclear matter, such as thermodynamic variables and transport coefficients~\cite{DXG2016-10,ZCL2013-11,SA2010-12,PGM1998-13}. All these observables contribute to the determination of the
equation of state (EOS)~\cite{gghz2014-14,blc2008-15} in nuclear matter and are also important in the description of the supernova core collapse and the formation of a neutron star~\cite{htj2007-16,pdr2002-17,jmm2004-18}. Among them, determination of temperature and shear viscosity over entropy density ratio $\left(\frac{\eta}{s}\right)$ are of very  interesting in heavy-ion collision. For the former, many efforts have been carried out by studies of various thermometers~\cite{SA2010-12,AJK2006-19,MCW2016-20}, and for the latter, it is found that the ratio of shear viscosity over entropy density of any fluids seems to have the bound of $\frac{1}{(4\pi)}$, proposed by Kovtun-Son-Starinets (KSS) in a certain supersymmetric gauge theory~\cite{PDA2005-21}. Many efforts to study the quark gluon plasma phase transition through these quantities have been performed~\cite{HSU2011-22,HGP2011-23,RNJ2007-24,RAJ2014-25,JYY2007-26}. For example, empirical observation of the temperature or incident energy dependence of the shear viscosity over entropy density reached its local minimum at the critical point for the phase transition~\cite{LJL2006-27}. In addition, there are many interesting studies on the ratio of $\frac{\eta}{s}$ in intermediate energy heavy-ion collisions~\cite{DXG2016-10,ZCL2013-11,SXL2011-28,SXL2011-29,FZ2016-30,DQF2014-31,SP2010-32,ZCL2012-33,ZCL2014-34,JLC2013-35,GCQ2017-36,jpo1995-6,NASS2009-38,DDN2011-39,LSPD2003-40}, especially focus on the nuclear liquid-gas phase transition by studying the behavior of $\frac{\eta}{s}$~\cite{DXG2016-10,ZCL2013-11,SP2010-32,ZCL2012-33,ZCL2014-34,JLC2013-35}.The intermediate energy nuclear reaction process is quite complex, where both the mean field effects and the two-body interactions play an equally important role and furthermore, the blocking effect is another important ingredient. The knowledge of the dissipation mechanism for nuclear matter in HIC is related to the properties of the mean-field via the one-body dissipation and nucleon-nucleon collisions via the two body dissipation in the nuclear medium. To treat the HIC dynamics in this energy range, two classes of transport theory models, namely Boltzmann-Uehling-Ulenbeck (BUU)~\cite{blc2008-15,GFSD1988-41,AFJ1994-42,OTK2012-43,YGM2013-44} model and Quantum molecular Dynamics (QMD) ~\cite{JA1991-45,RCJ1996-46} model, are very successful to describe various experimental observables. The former is a one-body transport theory and the latter is based on the many-body transport theory.

Many theoretical studies on the mean free path ($\lambda_{nn}$) and the in-medium nucleon-nucleon cross section ($\sigma_{nn}$) have been performed in the last decades~\cite{BS1983-47,ARVS2012-48,WUH1998-49,HFZ2007-50,WZJ2007-51,FS2008-52}. However, experimental results are difficult and not extensively reported so far. Recently, by looking at free protons specifically, the INDRA Collaboration presented a comprehensive body of experimental results concerning the mean free path, the nucleon-nucleon cross-section and the in-medium effects in nuclear matter by performing a systematic experimental study of nuclear stopping in central collisions for different combinations of heavy-ion induced reactions in the Fermi-energy domain below 100 MeV/u ~\cite{OL2014-53}. Nuclear stopping governs the amount of dissipated energy, the amplitude of large collective motion, and the competition between various mechanisms such as deep inelastic reaction, neck emission, and fusion reaction~\cite{GDO2010-54,GQZ2011-55,YZX2016-56}. It is found that $\lambda_{nn} = 9.5\underline{+}2$fm at $E_{beam} = 40$ MeV/u and $\lambda_{nn} = 4.5\underline{+}1$fm for $E_{beam} = 100 $MeV/u. In addition, they also estimated values of the in-medium nucleon-nucleon cross section when nuclear matter is in a compression state.

In this work, our motivation is to study the thermal and transport properties, and extract the information of the in-medium nucleon-nucleon cross section as well as the mean free path during the compression stage for the central collisions of $^{129}$Xe+$^{119}$Sn which is one of the focused systems in the INDRA data mentioned above~\cite{OL2014-53}, based on the isospin-dependent quantum molecular dynamics model (IQMD). Using the parametrization function proposed by Danilewiez~\cite{pdz1984-68,bwpd2010-69}, the time evolution of shear viscosity can be obtained. The in-medium nucleon-nucleon cross section is less than $10$mb below $100$ MeV/u in the central region of the reaction system during the compression stage, which is consistent with the experimental results from the INDRA data of Lopez et al.~\cite{OL2014-53}. The deduced mean free path is also consistent with the experimental result from the same data~\cite{OL2014-53} above $40$ MeV/u, which is close to the data value $4.5\underline{+}1$fm. Based on the reasonable simulation results of $\sigma_{nn}$ and $\lambda_{nn}$, we further deduce the shear viscosity over entropy density ratio for the hot nucleus formed in central collision $^{129}$Xe+$^{119}$Sn below 100 MeV/u, and find the $\frac{\eta}{s}$ value tends to $\frac{3}{4\pi}$, which indicates the hot nucleus behaves like a perfect fluid.

This paper is organized as follows: Section~\ref{modelformalisim} provides a brief introduction for the IQMD model and calculation formalisms. In Section~\ref{resultanddiscussion} we present the calculation results and discussion, where the time evolution of thermodynamic quantities and shear viscosity over entropy density ratio are focused. Finally a conclusion is given.

\section{MODEL AND FORMALISM}
\label{modelformalisim}
\subsection{MODEL}
The IQMD (Isospin-dependent Quantum Molecular Dynamics) model is an extended version of quantum molecular dynamics (QMD) model, which use a many-body theory to describe the behavior of heavy-ion collision at intermediate energies~\cite{JA1991-45,CLL1989-57,CH1998-58,TTW2015-59,ZQF2015-60,JC2016-61,WJX2015-62}. The IQMD model includes the effects of isospin and Pauli blocking. Each nucleon is represented by a Gaussian wave packet in the model, i.e.
\begin{equation}\label{wavefunction}
\phi_{i}\left(\mathbf{R},t\right) = \frac{1}{\left(2\pi L\right)^{3/4}}exp\left[-\frac{\left(\mathbf{R}-\mathbf{R}_{i}\right)^2}{4L}+\frac{i}{\hbar}\mathbf{r}\cdot\mathbf{p_{i}}\right],
\end{equation}
where $\mathbf{R}_{i}$ and $\mathbf{P}_{i}$ are the time dependent variables, respectively.  $L$ is the Gaussian width which is set to 2.16 $fm^{2}$ in present work . The total effective Hamiltonian can expressed as
\begin{equation}\label{hamidun}
\left<H\right> = \left<T\right>+\left<V\right>,
\end{equation}
where $\left<T\right>$ is the kinetic energy, and $\left<V\right>$ is the mean field part:
\begin{equation}\label{meanfield}
\left<V\right> = U_{Sky} + U_{Coul} + U_{Yuk} + U_{Sym},
\end{equation}
where $U_{Sky}$, $U_{Coul}$, $U_{Yuk}$, $U_{Sym}$ represent the Skyrme potential, the Coulomb potential, the Yukawa potential and the Symmetry potential, respectively. The Skyrme potential is
\begin{equation}\label{Skyrme}
U_{Sky} = \alpha\left(\frac{\rho}{\rho_{0}}\right)+\beta\left(\frac{\rho}{\rho_{0}}\right)^{\gamma}+
\delta*ln^{2}\left[\epsilon\left(\Delta\mathbf{P}\right)^{2}+1\right]\frac{\rho}{\rho_{i}},
\end{equation}
where $\rho$ is the total density of nuclear system, $\rho_{0}$ is the saturation density at ground state and $\rho_{0}$ = 0.168 $fm^{-3}$. $\alpha$, $\beta$, $\gamma$ are the Skyrme parameters, which connect with the EOS of bulk nuclear matter. $\delta$ and $\epsilon$ are parameters of the momentum-dependent potential~\cite{JA1991-45}. The symmetry potential is
\begin{equation}\label{symmetry}
U_{Sym}=\frac{C_{sym}}{2\rho_0}\sum_{i\not=j}\tau_{iz}\tau_{jz}\frac{1}{(4L\pi)^{3/2}}exp{-\frac{(\mathbf{r}_{i}-\mathbf{r}_{j})^2}{4L}}
\end{equation}
where $C_{sym}$ is the symmetry energy coefficient, $\tau_{z}$ is the zth component of the isospin degree of freedom for the nucleon, which equals 1 or -1 for nucleon or proton, respectively.
The expression of the other potentials can be found in Ref.~\cite{JA1991-45,CH1998-58}.
In present work, the parameter $\alpha$ = -390 MeV, $\beta$ = 320 MeV, $\gamma$ = 1.14, $\delta$ = 1.57 MeV and $\epsilon$ = 500 $\frac{c^2}{GeV^2}$ are used, corresponding to the soft EOS with momentum dependent interaction  and the incompressibility of $K$ = 200 MeV. With the present framework, reasonable phase-space information of nucleon in intermediate-energy collisions can be obtained.

\subsection{FORMULA}

Thermal properties of the hot nuclear matter can be extracted  in two steps. First, based on the IQMD simulation, one could calculate the nuclear matter densities at each point in coordinate space at every time step by
\begin{equation*}
\label{density}
\rho\left(\mathbf{R},t\right) = \sum_{i=1}^{A_{T}+A_{P}} \frac{1}{\left(2\pi L\right)^{3/2}}exp\frac{-\left(\mathbf{R}-\mathbf{R}_{i}\right)^2}{2L}
\end{equation*}
with the summation over all nucleons, where $L$ = 2.16 $fm^2$ as mentioned before. Second, using the value of nuclear matter density to calculate other thermodynamical properties of hot nuclear matter formed in heavy-ion collisions, such as temperature, shear viscosity, entropy density etc. In recent years, many thermometers to extract nuclear temperature have been developed, for instance, the slope temperature, double isotopic ratios and the population of excited states thermometer~\cite{JS2012-63,HZ2011-64,HZ2013-65,zhba2012-66}. Here, another method for measuring the temperature which was proposed in Ref.~\cite{SA2010-12} is applied, which is based on momentum fluctuations of detected particles. The variance is given by formula
\begin{equation}\label{variance}
\sigma_{xy}^{2} = \int\left(\left<Q_{xy}^{2}\right>-\left<Q_{xy}^{2}\right>^{2}\right)n\left(p\right)d^{3}p,
\end{equation}
where $n\left(p\right)$ is the momentum distribution of particles, the $Q_{xy}=p_{x}^2-p_{y}^2$ is the quadruple momentum, which is defined in a direction transverse to the beam axis and $p_{x}$, $p_{y}$ are the components of momentum vector extracted from the phase space of the IQMD model. The average in Eq.(~\ref{variance}) is performed over events. Proton and neutron, which are fermions and follow the Fermi statistics, carry important information on densities and temperatures in heavy-ion collision. Therefore the Fermi-Dirac distribution can be used in Eq.(~\ref{variance})~\cite{DQF2014-31,HZ2011-64,HZ2013-65,zhba2012-66,WB1995-67}. Using the Fermi-Dirac distribution $n\left(p\right)$, we can use
\begin{equation}
\begin{split}
\sigma_{xy}^{2}&=\frac{\int\left(p_{x}^{2}-p_{y}^{2}\right)^{2}n\left(p\right)\,d^{3}p}{\int n\left(p\right)\,d^{3}p}\\
&=\left(2mT\right)^{2}\frac{4}{15}\frac{\int_{0}^{\infty}y^{5/2}\frac{1}{exp\left(y-\upsilon\right)+1}\,dy}{\int_{0}^{\infty}y^{1/2}\frac{1}{exp\left(y-\upsilon\right)+1}\,dy}\\
&=\left(2mT\right)^{2}F_{QC}\left(z\right),
\end{split}
\end{equation}
where $F_{QC}(z)$ is the quantum correction factor and $z=exp({\mu/T})$, $\mu$ is the chemical potential of nucleon which is determined by the following implicit equation
\begin{equation}\label{chemical}
\frac{1}{2\pi^2}\left(\frac{2m}{\hbar^2}\right)^{3/2}\int_{0}^{\infty}\frac{\sqrt{\epsilon}}
{e^{\frac{\epsilon-\mu}{T}}+1}d\epsilon=\rho.
\end{equation}

Expanding to the lowest order in $\frac{T}{\varepsilon_{f}}$, where $\varepsilon_{f}=\varepsilon_{f0}\left(\frac{\rho}{\rho_{0}}\right)^{2/3} = 36\left(\frac{\rho}{\rho_{0}}\right)^{2/3}$ MeV is the Fermi energy of nuclear matter. The following result was obtained in Ref.~\cite{zhba2012-66}:
\begin{equation}\label{expanding}
\sigma_{xy}^{2} = \left(2mT\right)^{2}\frac{4}{35}\left[1+\frac{7}{6}\pi^{2}\left(\frac{T}{\varepsilon_{f}}\right)^{2}+o\left(\frac{T}{\varepsilon_{f}}\right)^{4}\right].
\end{equation}

The variance $\sigma_{xy}^{2}$, density and chemical potential can be calculated by phase space information from spatial distribution of nucleons, and with the Eq.(\ref{chemical})-Eq.(\ref{expanding}) the value of temperature of nuclear matter can be obtained.
In hydrodynamics, the internal friction occurs when there exists relative motions in a fluid (liquid and gas), and this is called viscosity. The nuclear shear viscosity has been derived  microscopically~\cite{,pdz1984-68,bwpd2010-69} and can be parameterizied as a function of density and temperature
\begin{equation}\small\label{shear}
\eta\left(\frac{\rho}{\rho_{0}},T\right) = \frac{1700}{T^{2}}\left(\frac{\rho}{\rho_{0}}\right)^{2}+\frac{22}{1+T^{-2}10^{-3}}\left(\frac{\rho}{\rho_{0}}\right)^{0.7}+\frac{5.8\sqrt{T}}{1+160T^{-2}},
\end{equation}
where $\eta$ is in MeV/$fm^2c$, T is MeV and $\rho_{0} = 0.168fm^{-3}$. In the kinetic theory~\cite{khsm1987-70,rpsm2003-71}, the shear viscosity ($\eta$) can be expressed as
\begin{equation}
\label{viscosity}
\eta = \frac{1}{3}m\rho\nu\lambda,
\end{equation}
where $\rho$ is the density, $\nu$ is the mean velocity of particles at given density and temperature, m is the mass of particle and $\lambda$ is the mean free path.
Assuming $\tilde{p} = m\nu$ is the mean momentum, the Eq.(\ref{viscosity}) can be expressed as
\begin{equation}\label{shearviscosity}
\eta = \frac{\tilde{p}}{3\sigma},
\end{equation}
where  $\sigma$ is the in-medium nucleon nucleon cross section. The relation of $\sigma$ and $\lambda$ is shown in following function:
\begin{equation}\label{lamuda}
\lambda = \frac{1}{\rho\sigma}.
\end{equation}

The mean momentum is a function of density and temperature. Using the nucleon Fermi-Dirac distribution, the mean momentum
\begin{equation}\label{meanmomentum}
\begin{split}
\tilde{p}\left(T,\rho\right)&=\frac{\int pn\left(p\right)d^3p}{\int n\left(p\right)d^3p}\\
&=\sqrt{2m}\frac{\int_{0}^{\infty} \frac{\epsilon}{e^{\frac{\epsilon-\mu}{T}}+1}d\epsilon}{\int_{0}^{\infty} \frac{\sqrt{\epsilon}}{e^{\frac{\epsilon-\mu}{T}}+1}d\epsilon},
\end{split}
\end{equation}
where $\epsilon = \frac{p^2}{2m}$ is the energy, $\mu$ is the chemical potential, $T$ is the temperature and $m$ is the mass of nucleon. So the mean momentum at a given density and temperature is obtained. With the shear viscosity from Eq.(\ref{shear}) and the mean momentum from Eq.(\ref{meanmomentum}), the in-medium nucleon-nucleon cross section
\begin{equation}\label{sigma}
\sigma = \frac{\tilde{p}}{3\eta}
\end{equation}
can be extracted. At the same time, the value of mean free path $\lambda$ is obtained, using the Eq.(\ref{lamuda}).

The nucleon-nucleon cross section in the IQMD model uses the cross-section of the experimental parametrization form which was also used by the VUU model, it includes elastic and inelastic channels, and distinguishes the isospin~\cite{fd1996-72,p2007-73,Ab2001-74}. In this context, we can label the nucleons and output the nucleon-nucleon cross section of each pair of nucleons at each time step.
In order to compare with the experimental results, we sum over the valid nucleon-nucleon (i.e. Pauli allowed) cross section of all nucleons and get the average value in the selected region as the in-medium nucleon-nucleon cross section which can be compared to the experimental result.
Then we use it and the value of shear viscosity calculated before to calculate the mean free path of reaction system by applying the Eq.(\ref{lamuda}).

The entropy density is calculated by
\begin{equation}\label{entropy}
s = \frac{U-A}{T}\cdot\frac{1}{V}=\left[\frac{5}{2}\frac{f_{5/2}\left(z\right)}{f_{3/2}\left(z\right)}-lnz\right]\rho,
\end{equation}
where $f_{m}\left(z\right) = \frac{1}{\Gamma\left(m\right)}\int_{0}^{\infty}\frac{x^{m-1}}{z^{-1}exp(x)+1}dx$, $U$ and $A$ are the internal and Holmholtz free energy, respectively~\cite{khsm1987-70,rpsm2003-71}.

\section{CALCULATION AND DISCUSSION}
\label{resultanddiscussion}

In this paper, central collisions of $^{129}$Xe + $^{119}$Sn at different beam energies from $15-100$ MeV/u are simulated by employing the IQMD model with the soft momentum-dependent interaction (MDI). In order to study the behavior of nucleons at the compressible stage, we choose the anterior $80$ fm/c as the research object during the heavy-ion collision. The central region is defined as a $\left[-3,3\right]^3fm^{3}$ cubic box and its center locates in the centre of mass of collision system. And in order to calculate quantities at the same condition, we chose the moment of two nuclei just touch as the zero moment of the time evolution.

\subsection{Time evolution of thermodynamic variables}

\begin{figure}
\includegraphics[width=16.4cm]{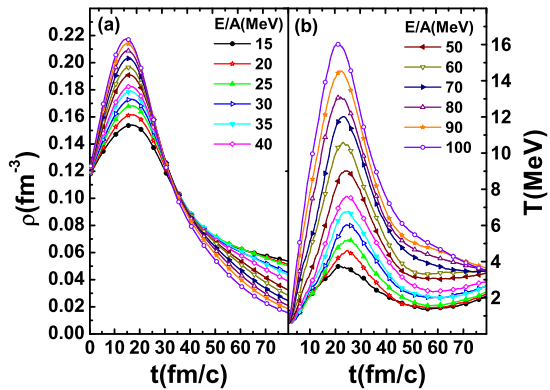}
\vspace{-6cm}
\caption{(Color online) The time evolution of the average density (a) and temperature (b) of central region at different incident energies.}
\label{denstemp}
\end{figure}

Fig.~\ref{denstemp} shows the time evolution of average density and temperature in the central region $\left[-3,3\right]^3fm^{3}$ at incident energies of 15-100 MeV/u, respectively. From Fig.~\ref{denstemp}$(a)$, it is seen that the average density increases in the compression stage  and decreases in the expansion stage. The average density reaches a maximum before $20$ fm/c. At  higher incident energy, the larger  average density is reached when the system at the highest compression.  Similar to the average density, the time evolution of temperature has the same trend at different energies, which is shown in Fig.~\ref{denstemp}$(b)$.
At a given beam energy, it is seen that at zero moment the initial temperature is close to zero. Then it increases to a local maximum around 20 fm/c and decreases at later times. The higher the incident energy, the larger the maximum value is. The corresponding time at maximum value of the temperature is a little later than that for the density, which was also observed in Ref.~\cite{DXG2016-10}, indicating that the momentum exchange among interacting nucleons does not cease even though the largest overlap of the system has been reached.

\begin{figure}
\includegraphics[width=16.4cm]{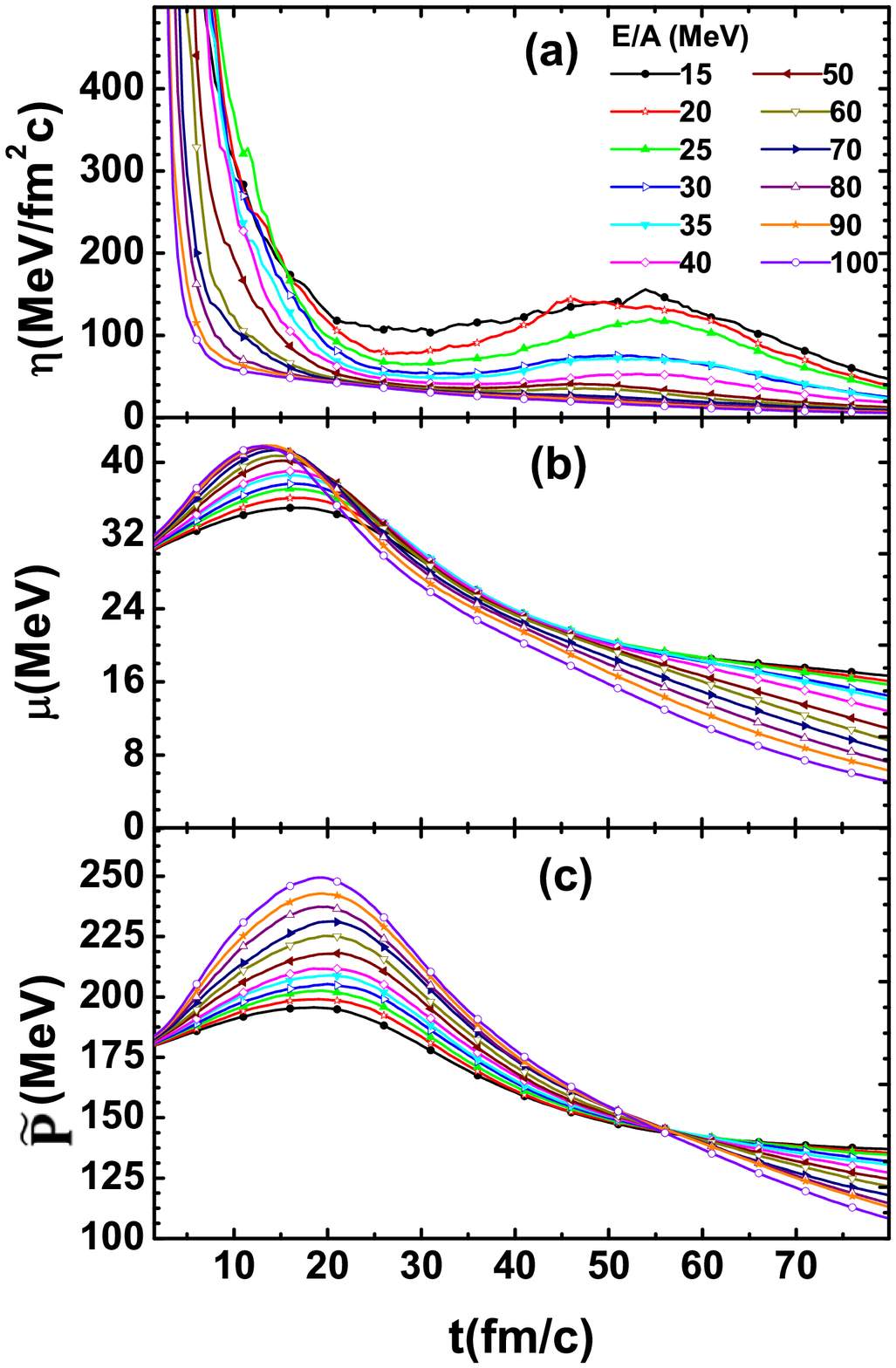}
\vspace{-14cm}
\caption{(Color online) The time evolution of shear viscosity (a), chemical potential (b) and mean momentum (c) in central region at different energies.}
\label{sheartt}
\end{figure}

The time evolution of shear viscosity, mean momentum and chemical potential are shown in Fig.\ref{sheartt}.
Here it should be noted that the formula of shear viscosity is principally applicable only when the
system is largely equilibrated. However, a full equilibrium is
hardly achieved during the whole heavy-ion collision process. Nevertheless, a local equilibrium could be more or less realized since the nuclear stopping characterizing the extent of equilibrium reaches its local maximum during the largest compression stage.
From top to bottom, Fig.~\ref{sheartt}(a) shows that in most cases $\eta$ drops at first $\sim$ 20 fm/c and tends to a local minimum around $\sim$ 25 fm/c. As Eq.(~\ref{shear}) shows, the shear viscosity depends on both temperature and density which vary with time. A combined effect leads to a local minimum of the shear viscosity.

\begin{figure}
\includegraphics[width=16.4cm]{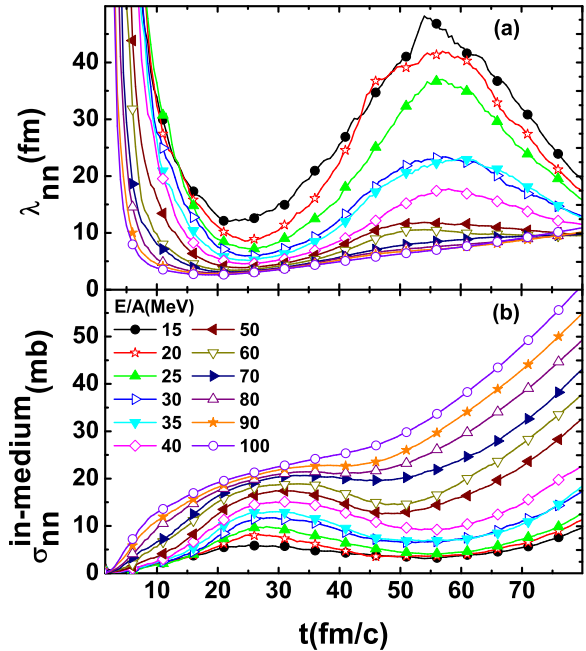}
\vspace{-10cm}
\caption{(Color online) The time evolution of  the mean free path (a) and the nucleon-nucleon cross section (b) of central region at different incident energies.}
\label{sigma-lamuda}
\end{figure}

Fig.\ref{sheartt}(b) shows the time evolution of the chemical potential ($\mu$). It is seen that $\mu$ increases in the compression stage and decreases in the expansion stage. Generally, its time evolution is very similar to the density evolution, i.e. the time reaching to their corresponding peaks seems very close to each other. With the chemical potential, density and temperature, the mean momentum of the central region is obtained, using the Eq.~(\ref{meanmomentum}). From Fig.~\ref{sheartt}(c), it can be seen that the time evolution of mean momentum has the same behavior as temperature at different energies, and the time of peak values are almost synchronous. Then, with Eq.~(\ref{sigma}) and Eq.~(\ref{lamuda}), the time evolution of the mean free path and the in-medium nucleon-nucleon cross section for the central region of the reaction system are shown in Fig.~\ref{sigma-lamuda}$(a)$ and Fig.\ref{sigma-lamuda}$(b)$, respectively. We can see that $\sigma_{nn}$ generally shows peaks around $\sim$ 30 fm/c which are a little delayed respect to the time of the peak of mean momentum due to a combination effects of the mean momentum and the shear viscosity, while the mean free path displays a wider valley around $20-25$ fm/c due to its inversely proportionality to $\rho\sigma_{nn}$. Roughly speaking, the mean free path tends to be the shortest during the compression stage due to the larger in-medium density and nucleon-nucleon cross section, which makes the nucleons transport in a short path range.

\begin{figure}
\includegraphics[width=16.4cm]{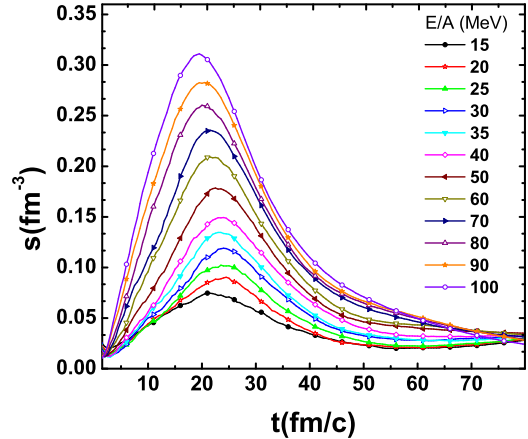}
\vspace{-6cm}
\caption{(Color online) The time evolution of entropy density in central region at different energies.}
\label{entr}
\end{figure}

The time evolution of entropy density is shown in Fig.~\ref{entr}. It is seen that the entropy density almost synchronically evolves with the temperature. At the compression stage, entropy density reaches a local maximum and decreases at expansion stage. The higher the incident energy, the larger the entropy density is.

\subsection{Excitation function of the thermodynamic variables at the highest compression}

\begin{figure}
\includegraphics[width=16.4cm]{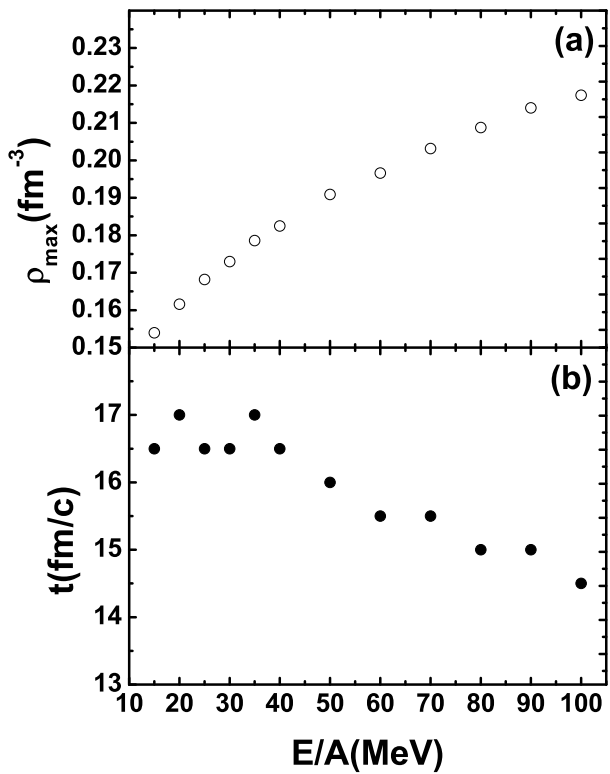}
\vspace{-13cm}
\caption{(Color online) The maximum density and the corresponding moment of central region at different energies.}
\label{maxdens}
\end{figure}

In the above figures, there exist maxima or minima of some quantities around  the largest compression stage. To compare the value of different quantities at the same time, we select the moment of maximum density at different energies as the given time, then the relationship between thermodynamic quantities with incident energy at the corresponding time are plotted. Fig.~\ref{maxdens} shows the energy dependence of maximum average density and the corresponding time at the maximum density in different beam energies. From Fig.~\ref{maxdens}(a), it illustrates that the maximum density can excess over normal density at $30$ MeV/u and it increases with the beam energy to $1.3\rho_{0}$ at 100 MeV/u. The corresponding time when the largest compression is reached decreases with the increase of the reaction evolution as depicted in Fig.\ref{maxdens}(b) due to the faster reaction kinematics.

\begin{figure}
\includegraphics[width=16.4cm]{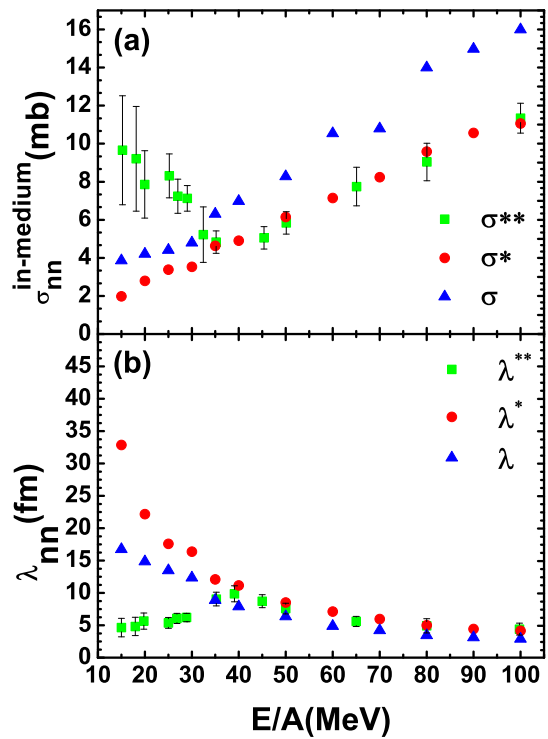}
\vspace{-12cm}
\caption{(Color online) (a) Energy dependence of in-medium nucleon-nucleon cross section. $\sigma^*$ and $\sigma$ are the result of the IQMD model output and the value calculated by formula, respectively. (b) Energy dependence of the mean free path. $\lambda^*$ and $\lambda$ are the value calculated from the value of $\sigma^*$ and $\sigma$ shown in Fig.~\ref{sigcompare}, respectively. $\sigma^{**}$ and $\lambda^{**}$ are the result from the Ref.~\cite{OL2014-53}.}
\label{sigcompare}
\end{figure}

\begin{figure}
\vspace{-1.cm}
\includegraphics[width=17.6cm]{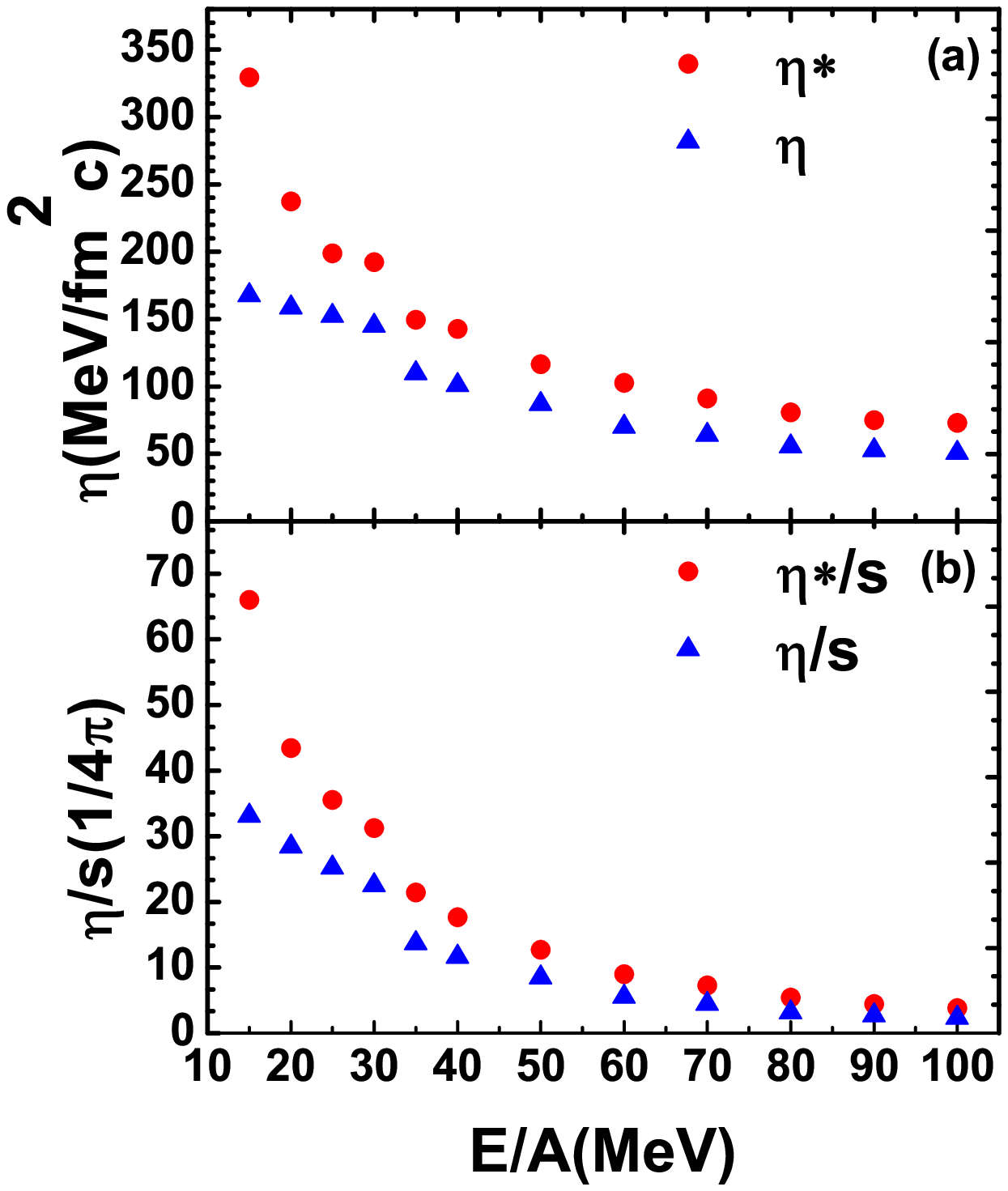}
\vspace{-12cm}
\caption{(Color online) Energy dependence of shear viscosity ($\eta$) (a) and the shear viscosity over entropy density ($\eta$/s, in units of $1/4\pi$) (b) at the largest compressible state. $\eta^*$ and $\eta$ are the values calculated from the value of $\sigma^*$ and the Eq.(\ref{viscosity}), respectively.
}
\label{eta_E}
\end{figure}

In order to study the behavior of the mean free path and the in-medium nucleon nucleon cross section at different incident energy in the largest compressed stage, their values are extracted at the corresponding time. The value of the in-medium nucleon-nucleon cross section of the IQMD model output is also obtained and using its value the mean free path is calculated employing the Eq.~(\ref{lamuda}). The result is shown in Fig.~\ref{sigcompare}(a) and Fig.~\ref{sigcompare}(b). We can see that the in-medium nucleon-nucleon cross section shows a general increasing behaviour with the beam energy regardless of the calculation methods ($\sigma^*$ is the IQMD output method and $\sigma$ comes from the Eq.~(\ref{sigma})). Even though the two results are quantitatively different, both  are  in a range of a few mb, which are close to the values of the in-medium nucleon-nucleon cross section($\sigma^{**}$) deduced from the experiments above 40 MeV/u~\cite{OL2014-53}. The calculated mean free paths  ($\lambda^*$ and $\lambda$)  have the similar trend as the experimental results when the energy is larger than 40 MeV/u in the maximum compressed stage and decrease toward an asymptotic value about 6 fm at $E$ = 100 MeV/u, which is very close to the deduced experimental  value  ($\lambda^{**}$) of Lopez {\it et al.} ~\cite{OL2014-53}.
 In addition, the shear viscosity can be obtained as displayed in Fig.~\ref{eta_E}. Similar to the behaviour of mean paths, the shear viscosity tends to decrease at higher energy and close to the similar values for both  calculations.

Since we have got reasonable results of the in-medium nucleon-nucleon cross section as well as mean free path, we can further deduce the shear viscosity and its ratio over entropy density at the largest compression stage, which are shown in Fig.~\ref{eta_E} for their energy dependence. Clearly, shear viscosity drops as a function of beam energy and trends to more or less saturated value around 100 MeV/u, indicating the hot nuclear matter becomes more fluid-like at higher beam energy. The ratio of shear viscosity over entropy density displays the similar decreases with the beam energy, which reaches to the value of around $\frac{2-4}{4\pi}$ at $80-100$ MeV/u. This asymptotic value is also consistent with our previous results from different models calculations for the hot nucleons~\cite{DXG2016-10,ZCL2013-11,SXL2011-28,DQF2014-31,ZCL2012-33,ZCL2014-34,JLC2013-35} as well as GDR~\cite{GCQ2017-36}, which means that this asymptotic value is less model dependent and demonstrates its fluid-like feature for hot nuclei formed in intermediate energy heavy ion collisions. In fact, this value is also not so different from a QCD matter which was found at RHIC and LHC where the matter is in the quark level, but here it is nucleonic matter.  Therefore,  the value of
$\eta$/s is not so much different even though it is very different microscopic matter level. Interestingly, it is noted that this $\eta$/s value is also close to the deduced experimental value from the GDR experimental data~\cite{DDS2017-75}.

\section{Summary and outlook}
\label{summary}
In summary, we studied thermodynamic variables and transport coefficients, specifically $\eta/s$  in central $^{129}$Xe + $^{119}$Sn collisions at intermediate energy in a framework of the isospin dependent quantum molecular dynamics  model. The properties of central region of nuclear reaction are discussed with some quantities. Time evolutions of density, temperature, entropy density and shear viscosity are presented, which give thermal and transport information on nuclear matter of collision system. Then using the kinetic theory, the mean free path and the in-medium nucleon-nucleon cross section are extracted. When the reaction system reaches the maximum compressed stage, the comparison between the values of the mean free path at different energies and the experimental results from Ref.~\cite{OL2014-53} is made, which shows a reasonable consistence with the data when the beam energy is above 40 MeV/u. Furthermore, the shear viscosity over entropy density is deduced at the highest compressible state in different energies, which demonstrates that a decreasing trend of $\eta$/s versus incident energy. When beam energy reaches to 100 MeV/u, $\eta$/s value is about three times of KSS bound, which indicates the hot nucleonic matter is very close to perfect fluid, not too much different from the partonic fluid, even though the particle level is very different. Since our simulated results of mean free path and in-medium nucleon-nucleon cross section are very close to the data, we believe that our derived shear viscosity over entropy density could be regarded as the experimental one.

\begin{acknowledgments}

This work is supported in part by the National Natural Science Foundation of China under Contracts Nos. 11421505 and 11220101005, the Major State Basic Research Development Program in China under Contracts No. 2014CB845401, the Key Research Program of Frontier Sciences of CAS under Grant  No.
QYZDJSSW-SLH002, the Strategic Priority Research Program of the Chinese Academy of Sciences under Grant No. XDB16, Chinese Academy of Sciences (CAS) President$'$s International Fellowship Initiative No. 2011T2J13 (M.V.) and 2015VWA070 (A.B.), the Slovak Scientific Grant Agency under contract
2/0129/17, and the Slovak Research and Development Agency under contract APVV-0177-11 (M.V.).

\end{acknowledgments}

\end{CJK*}
\end{document}